\documentstyle[aps,psfig,multicol]{revtex}
\begin{document}

\title{Controlling extended systems with spatially filtered, 
time-delayed feedback}
\author{M.~E.~Bleich~$^{a}$, D.~Hochheiser~$^{b}$, 
J.~V.~Moloney~$^{b}$,
 and J.~E.~S.~Socolar~$^{a}$}

\address {$^{a}$~Department of Physics and Center for Nonlinear 
	and Complex Systems, Duke University, Durham, NC 27708\\
	$^{b}$~Arizona Center for Mathematical Sciences, 
		Department of Mathematics, University of Arizona, 
		Tucson, AZ 85721}
\maketitle

\begin{abstract}

We investigate a control technique for spatially extended systems
combining spatial filtering with a 
previously studied form of time-delay feedback.
The scheme is naturally suited to real-time control 
of  optical systems.
We apply the control scheme to a model of a transversely extended
semiconductor laser in which a desirable, 
coherent traveling wave state exists, 
but is a member of a nowhere stable family.
Our scheme stabilizes this state, and
directs the system towards it from realistic, distant
and noisy initial conditions.
As confirmed by numerical simulation, 
a linear stability analysis about the controlled state
accurately predicts when the scheme is successful,
and illustrates some key features of the control
including the individual merit of, and interplay between,
the spatial and temporal degrees of freedom in the control.

\end{abstract}


\begin{multicols}{2}

\narrowtext

\section{introduction}

Nonlinear dynamical systems often possess periodic orbits
that have desirable properties but are unstable.
The problem of applying small perturbations to the system
in such a way as to produce stable periodic behavior has
received much attention recently. \cite{ccreview}
This paper addresses the control problem as it arises in 
a specific context: the stabilization of
unstable traveling wave states of spatially extended systems.
Though such states have a particularly simple structure,
the control problem is nontrivial.

The general method introduced below may be applicable
to a wide variety of physical systems, but an entirely
general analysis of it is beyond the scope of this work.
We have chosen to investigate in detail its application to
two sets of model equations describing the dynamics of
wide aperture semiconductor lasers. 
Our results demonstrate that
unstable traveling wave states can be effectively controlled
in these systems
and therefore have implications both for the
general theory of control of spatially extended systems
and for the design of semiconductor lasers.

For the purposes of this paper, controlling a system
means providing feedback that
locks the system to one member of a possibly infinite 
family of unstable periodic orbits
present in that system, thereby choosing a desired  state
from a large variety of possibilities.
The technological goal is to produce a desirable
behavior in a system by applying carefully chosen feedback
that directs the system to the goal state and keeps it there.
For many applications, it is desirable to design the feedback
such that the magnitude of the control signal 
decreases as the system approaches the desired state, and,
in the absence of noise,
vanishes when the controlled behavior is a 
dynamical state of the uncontrolled system.
It is also worthwhile to consider ``controlling'' a
state which is only approximately a true orbit of the system. 
An important example is the situation where a stress is applied
to a large but finite transverse portion of a system.
Useful results for the dynamics of the active region
may still be obtained by using feedback
designed for the traveling wave solution of the infinite, 
idealized system.
In this case one might expect the feedback to become small, 
but not completely vanish.

For a system with an accessible dynamical field $A(x,t)$,
our control signal $\epsilon_A({\bf x},t)$ 
is derived from an infinite sum of signals
delayed in time by integer multiples of the period of the
state that is to be stabilized:
\begin{eqnarray} \label{controlterm}
&\epsilon_A({\bf x},t) = &   \\
&  &\gamma \left[ A({\bf x},t) - 
(1-R)\sum^\infty_{n=1}R^{n-1}{\tilde A_n}({\bf x},t-n\tau) \right], \nonumber
\end{eqnarray}
where 
$\gamma$ is the gain of the feedback, 
$\tau$ is the period of the target state, and 
$0 \leq R <1$ determines the relative weight
given to states farther in the past.
The field 
${\tilde A_{n}(x,t)} = {\cal F}^{-1} [f^{2n}(k-k_c) 
{\cal F}[ A(x,t)]]$ 
is the spatially filtered version of $A$.
Here $f()$ is a filtering function applied in Fourier space and
${\cal F}[A]$ is the spatial Fourier transform of $A$.
The precise manner in which the spatial filtering is 
included may vary;
we have made a choice that corresponds directly to 
an experimental arrangement described below.
We take $f(q)$ to be peaked around zero so that
contributions to $A$ from wave numbers other than the desired
wave number, $k_c$, are suppressed in ${\tilde A}$.
We also take $f(0)=1$ so that the feedback 
term vanishes identically
when the system is in a pure state of wave number $k_c$ that is
an unstable orbit of the uncontrolled system.
Eq.~\ref{controlterm}  represents the enhancement of 
time-delay feedback 
of the form analyzed in Ref.~\cite{BS}
with spatial filtering of the type introduced in Ref.~\cite{HLM}.

Control based on Eq.~\ref{controlterm}
is especially well-suited to spatially extended states
with a structure dominated by one Fourier mode.
Feedback occurs whenever there are components in
the system due to undesired wave numbers or undesired frequencies.
The temporal feedback is important both because
practical implementations
of the spatially filtered feedback necessarily involve
time delays and because spatially filtered feedback alone
is sometimes not sufficient to stabilize the desired state.

We emphasize that the control scheme investigated here is a 
plausible candidate for implementation in experimental systems.
As in Ref.~\cite{BS}, a form of time-delay feedback
is used which includes previous comparisons made between the state 
and a time-delayed version in a way that is easy to 
implement because the
feedback signal can be generated recursively. \cite{recursive}
This is a generalization of a low-dimensional 
control technique known as 
extended time-delay autosychronization, 
\cite{etdas} which is in turn
a generalization of a scheme introduced by Pyragas \cite{pyragas}
and has been demonstrated to work 
in low-dimensional electronic systems. \cite{etdas}

Optical systems are of particular interest with respect to control
both because they offer excellent laboratories for testing
theoretical ideas and because important technological problems 
associated with them  may be solved through control techniques.  
Wide aperture semiconductor lasers, with their compact 
size and very large gain, are ideal candidates for 
high brightness coherent
steerable laser sources (spatially and temporally coherent). 
However, the pronounced 
asymmetry in their gain and refractive index 
spectra leads to a very 
strong nonlinear amplitude-phase coupling in the laser field. 
Consequently, wide aperture semiconductor lasers display 
uncontrolled dynamic intensity filamentation 
(random beam steering) 
even immediately beyond lasing threshold. 
This behavior persists and becomes even more complicated at 
higher current pumping levels. 
Moreover the timescales involved in the laser dynamics lie 
in the nanosecond to picosecond regime, ruling out any algorithm 
which requires indirect intervention 
in order to establish control. 
Time delay feedback with real-time filtering in 
space and time is a  
natural candidate for all-optical control of these systems.

To illustrate the power of spatially filtered, 
time delay feedback,
we analyze the important example of the
laser Swift-Hohenberg equations 
which approximately describe the dynamics
of the optical field of a wide aperture semiconductor laser
with one transverse dimension.
Results of linear stability analyses are used to guide
the choice of control parameters for numerical simulations,
which reveal that the controlled state may be attained
even when the initial conditions are far from the
linear regime.
Also, because semiconductor edge-emitting lasers 
typically run on multiple longitudinal modes,
we study the case
of a two-longitudinal mode laser Swift-Hohenberg model.
We find that the presence of a second mode introduces
new features relevant to the control scheme, but that
control is still possible.

\begin{figure}[t]
\centerline{\hbox{
\psfig{figure=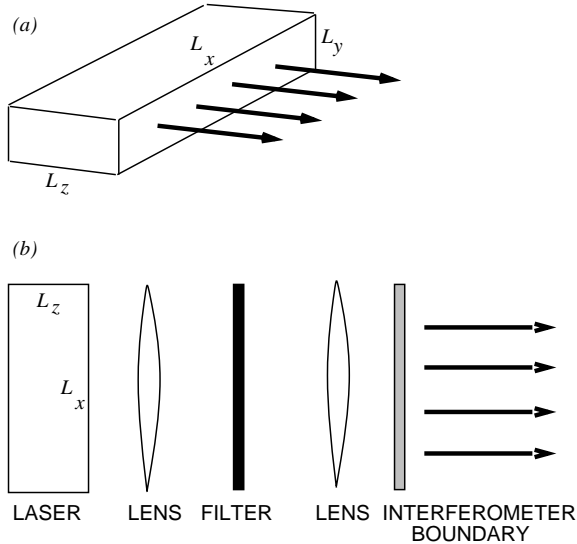,width=3in}
}}
 \caption{(a) Geometry of a wide aperture laser.  $L_y$ is 
 	assumed to be
	sufficiently small that only one mode dominates in 
	the $y$-direction.
	The large width $L_x$ gives rise to many transverse modes.
	The linear stability analysis in this paper is 
	valid for arbitrarily
	large $L_x$.  The size of $L_z$ determines the 
	number of longitudinal
	modes that are relevant for the dynamics.
 	(b) One possible schematic implementation of 
 	the feedback mechanism
	studied in this paper.  The feedback 
	signal is the field reflected
	from the front of the laser cavity in 
	the presence of an additional
	reflecting interface 
	(labeled ``Interferometer Boundary'').
	The spatial filtering is performed by the 
	two lenses with a filter
	placed at the focal plane of each.
         }
 \label{fsetup}	
\end{figure}

Our primary motivation for studying this particular system
is that the feedback signal of interest can be produced 
using a Fabry-Perot interferometer containing a spatial filter.
The time delay in the feedback scheme corresponds to the round
trip transit time in the interferometer.
The spatial filtering can
be accomplished by a focusing lens, 
whose focal plane contains the 
far-field fluctuating output of the laser. 
Placing a suitable aperture in the focal 
plane of the lens acts as a 
narrow band spatial filter. 
One example of a suitable arrangement 
for generating the desired feedback is shown in Fig.~\ref{fsetup}.
Our results therefore suggest a feasible approach to the
suppression of unwanted spatiotemporal fluctuations in
real laser systems.

One of the main results of the present paper 
is that the addition of spatial filtering
to time-delay control 
(using ${\tilde A_{n}}$ rather than $A_n$ in Eqn.~\ref{controlterm})
produces a highly robust control scheme.
In the context of the laser equations discussed below,
linear stability analysis of the infinite system
shows that in general both the spatial filtering and
the time delay are important components of the scheme.
For the models we study, 
numerical results also show that
when a state is stable with feedback
it is also highly attracting, 
so that linear stability analysis is predictive
even far from the linear regime.
Simulations of the model equations show that the feedback
is able to direct the system towards the desired state
from a distant initial condition,
and that spatial filtering is the dominant mechanism 
responsible for this behavior.

Before proceeding to the detailed analysis of our scheme,
we mention some related investigations.
First, results of numerical simulations of the 
application of our scheme with $R=0$ to
the control of traveling wave solutions
in the single longitudinal mode 
semiconductor laser Swift-Hohenberg
equations (discussed below), 
have been reported elsewhere. \cite{HLM}
There, the spatial filtering was shown to be extremely robust, 
rapidly suppressing the broadband noisy spatial spectrum of the 
free running unstable laser, in favor of the filtered transverse 
traveling wave mode. 
The subsequent evolution towards a controlled 
state was observed to depend sensitively on whether the system is 
infinitely  extended (idealized) or pumped over a finite transverse 
cross-section. 
In the former case, the system evolves
to a pure nonlinear traveling wave mode of 
the isolated laser system, 
although at higher stress the system 
typically spent time in a metastable
dynamical state before reaching the desired traveling wave.
For finite transverse pumping, 
the system evolves to a mixed traveling 
wave solution of the joint 
laser-feedback system and remains there. 
In this latter case, a finite 
amount of energy remains in the feedback loop.    

Second, Lu et al. \cite{Harrison} 
considered feedback constructed by
combining comparisons of the current state of the system 
both to a version delayed by the temporal 
period of the target state
and to versions shifted by the spatial period of the target state.
Numerical integration showed that the control can direct the system
towards, and stabilize, a pattern in a 
transversely extended laser model,
but the method does not appear to be a 
good candidate for all-optical
implementation.

Finally, Battogtokh et al. \cite{BM} considered the effect of
feeding back a time-delayed signal constructed from the global
average of a dynamical field, 
showing that stable uniform oscillatory
states of the system with feedback exist for some choices of
the delay time.
In the scheme they analyze, the control signal does not represent
a small perturbation and the delay time is not tuned to
the period of the desired orbit.

\section{Single longitudinal mode laser swift-hohenberg equations}

We now treat the specific example of 
a recently derived model of the transversely 
extended semiconductor laser, 
the semiconductor laser Swift-Hohenberg equations, \cite{Bowman}
extending the results of Ref.~\cite{HLM}.
The model assumes the cavity geometry shown in Fig.~\ref{fsetup}(a)
with $L_y$ and $L_z$ both small enough 
that the dynamics is dominated
by a single mode in the $y$ and $z$ directions, but $L_x$ large.
Denoting the $x$-dependent envelope of the electric field 
by the complex field $\psi$
and the carrier density by the real field $n$, the equations are
\begin{eqnarray} 
(\sigma + 1) \partial_t \psi &  = 
	& \sigma (r-1) \psi + i a \nabla^2 \psi
	- i \sigma \Omega \psi \nonumber \\ 
    & &  - \frac{\sigma}{(1+\sigma)^2}(\Omega+a \nabla^2)^2 \psi \\
    & & - \sigma (1+i \alpha) n \psi + 
	\epsilon_\psi \label{esmpsi} \nonumber \\
\partial_t n & = & -b n +  |\psi| ^2 \label{esmn}
\end{eqnarray}
where $\sigma$ and $b$ are the decay rates of the electric
field and population inversion respectively, 
normalized to the decay rate of the polarization,
$r$ is the scaled pump rate,
$a$ is a scaled diffusion constant,
$\Omega$ is the detuning between the atomic and carrier frequencies,
and $\alpha < 0$ is a nonlinear amplitude-phase coupling.
(All the coefficients in the equation are real.)

The model is similar to the laser 
Swift-Hohenberg equations \cite{LMN2}
for two-level lasers. 
The key difference in the semiconductor equations is
the explicit inclusion of the $\alpha$ term
which derives from the strong assymmetry in the 
semiconductor optical gain and 
refractive index spectra \cite{Henry,Chow}. 
Other terms arising in the semiconductor
version due to spectral hole burning in the carrier distributions
do not influence the results discussed here.

With control turned off ($\epsilon_{\psi}=0$),
Eqns.~(\ref{esmpsi}) and~(\ref{esmn}) have traveling wave solutions 
that are always unstable \cite{LMN2}:
\begin{eqnarray}
\psi_k(x,t) & = & \rho \exp i(kx - \omega t + \phi), \\
n_k & = & \rho^2/b,
\end{eqnarray}
where $\phi$ is an arbitrary phase that will henceforth 
be assumed to be zero,
\begin{eqnarray} \
\rho^2 & = & b \left[ r - 1 - \left(\frac{\Omega - 
a k^2}{1+\sigma}\right)^2 \right],\label{rho}\\
\omega & = & \frac{\sigma \Omega + a k^2 + \alpha 
\sigma \rho^2/b}{1+\sigma}.
\end{eqnarray}
Note that $\rho$ is real and that the traveling wave solution
ceases to exist when the right-hand side of 
Eqn.~(\ref{rho}) is negative.

Eqn.~(\ref{esmpsi}) for the envelope of the electric field contains
a time-delay and spatial filtering control term of the 
form of Eqn.~(\ref{controlterm}),
with the time delay $\tau$ set to $2 \pi/\omega$, 
the period of the desired 
traveling wave state.
The insertion of the feedback as simply an additive term in
this equation is an approximation of the 
real effect of the feedback,
which actually consists of an electric 
field applied at the front face of
the laser cavity due to reflections 
from the elements shown in Fig.~\ref{fsetup}(b).
The gain $\gamma$ should be thought of as a phenomenological
parameter that characterizes the effect of this boundary term
on the longitudinal mode in question.
The optimal choice of the filter function 
$f$ is not immediately clear.
For now we take $f(q)$ to be a gaussian of width $\Gamma$; i.e.,
$f(q) = \exp[-q^2/\Gamma^2]$.
As explained below, 
the results are not sensitive to the precise choice of $\Gamma$.
The case of a square filter function is 
also discussed briefly below.

To perform the linear stability analysis we write
$\psi(x,t) = [1+B(x,t)] \psi_k$ and $n(x,t)=[1+C(x,t)] n_k$,
and arrive at the following linearized
equations in the vicinity of a traveling wave solution:
\begin{eqnarray}
(\sigma+1) \partial_t B & = & -(2 a k+4 i a k 
	{\tilde \sigma} \Omega
	  - 4 i a^2 {\tilde \sigma} k^3) \nabla B \nonumber \\
   & &   + (i a - 2 a {\tilde \sigma} \Omega 
	  + 6 {\tilde \sigma} a^2 k^2) \nabla^2 B \nonumber \\
   & & - 4 i {\tilde \sigma} a^2 k \nabla^3 B - 
   {\tilde \sigma} a^2 \nabla^4 B \nonumber \\
   & &  - (1+i \alpha) \frac{R^2}{b} \sigma C  + 
	 \epsilon_B \label{linearB} \\
\partial_t C & = & b (B + B^* - C) \label{linearC}, 
\end{eqnarray}
where ${\tilde \sigma} \equiv \sigma / (1+\sigma)^2$.

\begin{figure}
\centerline{\hbox{
\psfig{figure=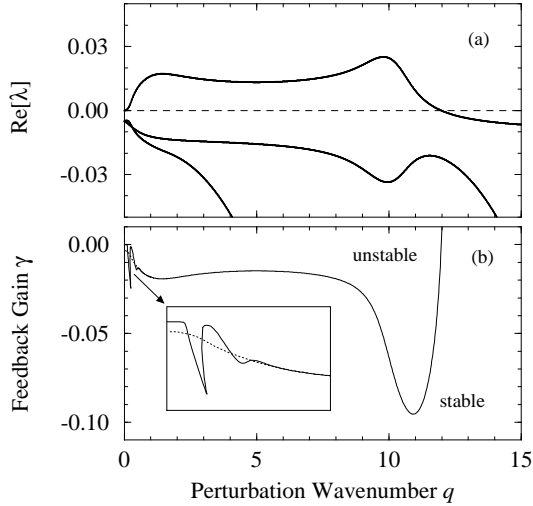,width=3.5in}
}}
 \caption{(a) Growth rates of perturbations of an 
 	uncontrolled ($\gamma=0$)
	traveling wave solution of 
	Eqns.~(8,9) 
	with $k=5$ and $r=1.5$.	
	and parameter values
	$\sigma=.1$, $\Omega=.001$, $a=.01$, 
	$b=.01$, and $\alpha=-5$.
	For each $q$, ${\bf Re[\lambda]}>0$ 
	implies exponential growth of the perturbation.
 	(b) Stable region in the $q,\gamma$ 
 	plane of the same traveling wave with control.
	The solid line is the stability boundary for $R=0$, 
	$\tau=2 \pi/\omega$,
	and $\Gamma = .25$.
	Note that all modes are stable for $\gamma < -0.1$.
	The dashed line (see inset) corresponds to the 
	(unphysical) case of 
	$\tau=0$, $R=0$. 
	}
 \label{fsmgrowth}	
\end{figure}

Following Ref.~\cite{BS}, we 
obtain a linear system of ordinary differential equations for the
Fourier modes of the perturbation.
Letting ${\bf\xi} = (B_q,B^*_q,C_q)$, the 3D vector 
of Fourier amplitudes
at wave number $q$, the equations can be written 
in the general form
\begin{equation} \label{general}
\frac{d}{dt}{\bf\xi} = {\bf J} \cdot {\bf\xi} + 
{\bf M} \cdot \epsilon_{\bf\xi},
\end{equation}
where $\epsilon$ is given by the expression in 
Eqn.~(\ref{controlterm}),
${\bf J}$ is obtained from the coefficients of 
Eqns.~(\ref{linearB}) and~(\ref{linearC}), 
and ${\bf M}$ is determined by which variables 
form the control signal
and how the control signal enters the equations.
In the present case, 
\begin{equation}
 {\bf M} = \left(\begin{array}{ccc} 
			1 & 0 & 0 \\
			0 & 1 & 0 \\
			0 & 0 & 0 
		\end{array}\right).
\end{equation}

The factor by which a given eigenmode of the perturbation grows
during one period of the evolution of the controlled system is
called a Floquet multiplier.
The time delay in the control term requires that
the initial conditions for the evolution must specify
the behavior over an entire continuous time interval of one period, 
so each spatial Fourier mode has an infinite number of 
eigenmodes and
Floquet multipliers, $\mu_j$.
Letting ${\bf\xi}^{(j)}$ represent the $j^{th}$ eigenmode, 
we have, by definition,
\begin{equation} \label{floquetmult}
{\bf\xi}^{(j)}(t+\tau) = \mu_j {\bf\xi}^{(j)}(t).
\end{equation}
The set of Floquet multipliers for a perturbation with wave number
$q$ determine that perturbation's linear stability;  
if one or more
multiplier has $|\mu_j| > 1$, the perturbation is unstable.

\begin{figure}
\centerline{\hbox{
\psfig{figure=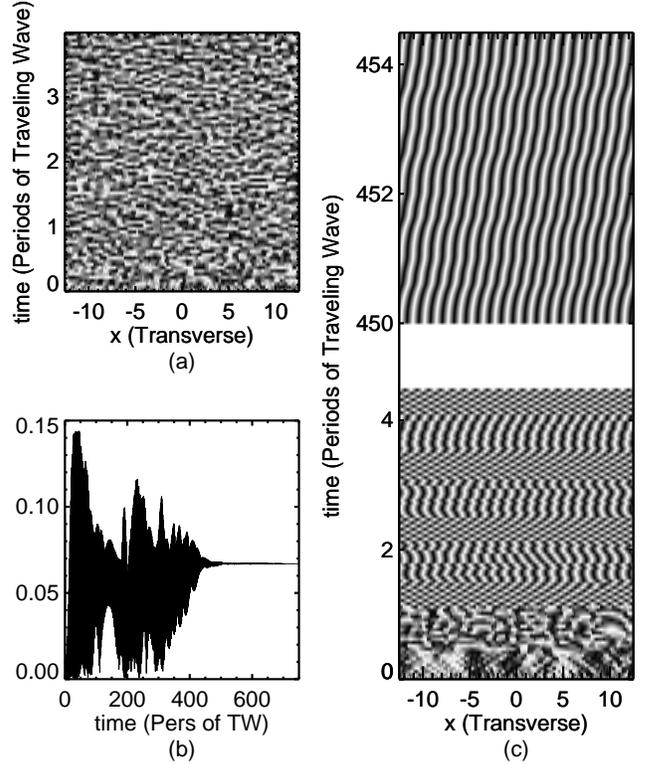,width=3.5in}
}}
 \caption{Evolution of the single mode system with $a, b, 
 	\Omega, \sigma,
	\alpha, r, k$ as in 
	Fig.~2a.
	, and $\gamma = .2$.
	(a) A spacetime plot of the phase of the field $\psi$ in 
	the uncontrolled system showing chaotic fluctuations.
	(b) The growth of the magnitude of the desired mode 
	as a function of time.
	(c) A spacetime plot of the phase of $\psi$ in the 
	controlled system
	for the same run as shown in (b).
	The lower region shows the dynamics when the system 
	is first turned on.
	After a short period during which fluctuations grow 
	rapidly, the feedback
	suppresses all wave numbers other than the desired one.  
	After a transient
	time of approximately 500 periods of the desired orbit, 
	the system settles
	into the traveling wave state.  The upper portion shows 
	the system as
	as it approaches the desired state, which would appear 
	as straight bands
	on this plot.
	}
 \label{fsmevolution}
\end{figure}

Dropping the subscript $j$ and evaluating 
the geometric sum in $\epsilon$,
Eqn.~(\ref{general}) may be written
\begin{equation}
\frac{d}{dt}{\bf\xi} = {\bf J} \cdot {\bf\xi} + 
	\gamma \left(
	\frac{1-f^{2}(q)\mu^{-1}}{1-R f^{2}(q)\mu^{-1}} \right) 
	{\bf M} \cdot {\bf\xi}.
\end{equation}
The values of the Floquet multipliers are determined by
requiring consistency between this equation and 
the defining relation
of Eqn.~(\ref{floquetmult}).
We obtain the following characteristic equation for this modified
eigenvalue problem:
\begin{equation}
g(\mu^{-1}) = \left| \mu^{-1} e^{\tau \left[{\bf J} 
+ \gamma \left(
	\frac{1-f^2(q)\mu^{-1}}{1-R f^2(q)\mu^{-1}} 
	\right) {\bf M} \right]} 
	- 1\!\!1 \right| = 0,
\label{def}
\end{equation}
where the exponential represents 
the operator that advances the linear system
by one period $\tau$.
As discussed in Refs.~\cite{BSold,BS}, one can perform
a numerical winding number calculation of $g(\mu^{-1})$
around the unit circle to obtain the number of
roots satisfying $|\mu^{-1}|<1$.
Since there are no poles in the unit disk,
the system is linearly stable if and only if this 
winding number vanishes.

Results from the linear stability analysis predict that our control
technique successfully stabilizes all traveling wave
solutions in the single longitudinal mode model.
We present detailed results for a single traveling wave solution,
for $k=5$ and $r=1.5$,
which is typical of all traveling waves we have studied.
(Values of the other parameters are given in the caption.)

Fig.~\ref{fsmgrowth}a shows the growth rates 
of the modes of the uncontrolled system, 
which are obtained by finding the eigenvalues of ${\bf J}$
in Eqn.~\ref{general}.  
There is one unstable mode for perturbation
wave numbers between zero and $\sim 12$.
With control, using $R=0$, 
we find that the traveling wave state is stable for 
$\gamma$ sufficiently negative.  
The solid line in Fig.~\ref{fsmgrowth}b 
indicates the boundary between which perturbation 
wave numbers are stable
or unstable at a given $\gamma$.
The controlled traveling wave is stable at values
of $\gamma$ for which all wave numbers are stable,
i.e., where the shaded region contains an entire horizontal line.
For all traveling waves in this model,
there is a minimum $|\gamma|$ for which the state is stable.
In the case shown in Fig.~\ref{fsmgrowth}b, 
this occurs at $\gamma \sim -.1$.
In this model, there is no lower boundary to the stable
region for traveling waves.

We find for this system that
spatial filtering alone would be sufficient to stabilize
the traveling wave.
The stability boundary obtained with
$\tau=0$ and $R=0$, the dashed line in Fig.~\ref{fsmgrowth}b,
is nearly identical to the one with $\tau=2 \pi/\omega$
at large $q$, but the time delay clearly has a significant effect
at $q$ near zero.
It is also important to note that implementation of
a spatial filter with no time delay is not possible
in fast optical systems.
The result that the introduction of a time delay
of one period does not {\em destroy} the stability
in the case of the gaussian filter is therefore significant.

In general, a given wave number perturbation can be stabilized
{\em either} by the time delay feedback with no spatial filtering
{\em or} by the spatial filtering with no time delay.
In each case, however, there may be small bands of wave numbers
for which one or the other method fails.
A given spatial filter fails near $q=0$ if there exist
perturbations which are sufficiently unstable
(or if $f(q)$ is sufficiently close to unity).
For $q$'s at which $f(q)=1$, 
as occurs for a finite band in the step function case,
the spatial filtering has no effect on the stability.
The time delay feedback alone fails for wave numbers whose
frequency of oscillation is sufficiently close to
an integer multiple of the frequency of the desired traveling wave.
Combining the spatial filter and the time delay renders
the system stable at all $q$.

The time delay is a crucial component for stability in the
two mode system discussed below.
In the single mode system, it may also play an important role
if $f(q)$ is chosen to be a step function
rather than a gaussian.

The predictions of several stability diagrams similar 
to Fig.~\ref{fsmgrowth}b
have been checked in detail by numerical simulation.  
The numerics show that the traveling wave states 
are stabilized with values of
$\gamma$ predicted by the linear analysis, 
and that instabilities occur at the wave number predicted 
when $|\gamma|$ is too small.

An important question is whether the linear stability analysis
is predictive of the behavior of the system even for
initial conditions that are not in the linear regime.
Numerical integration of the model equations show
that the spatially filtered feedback is particularly effective in
directing the system to the desired state.
As illustrated in Fig.~\ref{fsmevolution}, 
for parameters corresponding to a linearly stable controlled state,
the system is attracted to the desired state from
a typical initial condition.
Though it is difficult to display the full behavior during the long
transient, an investigation of the details reveals that, 
beginning from low amplitude noise
of the type that would be expected when the laser is 
first turned on,
the system, depending on the parameter regime,
may pass through several nearly stable states with the
desired wave number, but the incorrect frequency,
before finally settling on the one with the desired frequency.
Preliminary investigations of systems with time delay feedback
alone indicate that more complicated behavior occurs beyond
the linear regime.
  
\section{Two longitudinal mode laser swift-hohenberg equations}

Semiconductor lasers of practical interest generally operate
in regimes where many longitudinal modes may be active.
To begin to understand the possible effects of multiple 
longitudinal modes,
we study a two mode model.
This model is a straightforward generalization of the two level, 
one mode model
derived in Ref.~\cite{LMN2} to the situation in which 
two longitudinal
modes, with mode separation $2 \Delta$,
dominate the dynamics. \cite{modesep}
With the addition of the semiconductor 
$\alpha$ term discussed above,
this model reads,
\begin{eqnarray}
(\sigma + 1) \partial_t \psi_1 & = & \sigma (r-1) \psi_1 +
	i a \nabla^2 \psi_1 + i \Delta \psi_1 - i \sigma \Omega \psi_1 \nonumber \\
     & &  - \frac{\sigma}{(1+\sigma)^2} 
     (\Omega + \Delta + a \nabla^2)^2 \psi_1 \nonumber \\
     & & - \sigma (1+i \alpha) n \psi_1 - 
	\sigma (1+i \alpha) \eta \psi_2  + 
	\epsilon_{\psi} \label{etmpsi1} \\
(\sigma + 1) \partial_t \psi_2 & = & \sigma (r-1) \psi_2 +
	i a \nabla^2 \psi_2 - i \Delta \psi_2 - 
	i \sigma \Omega \psi_2 \nonumber \\
     & & - \frac{\sigma}{(1+\sigma)^2} (\Omega - 
     \Delta + a \nabla^2)^2 \psi_2 \nonumber \\
    & & - \sigma (1+i \alpha) n \psi_2 
    - \sigma (1+i \alpha) \eta^* \psi_1 + 
	\epsilon_{\psi} \label{etmpsi2} \\
\partial_t n & = & -b n + |\psi_1|^2 + |\psi_2|^2 \label{etmn} \\
\partial_t \eta & = & -b \eta + \psi_1 \psi_2^*.  \label{etmeta}
\end{eqnarray}
Note that the same control term 
$\epsilon_{\psi}\equiv\epsilon_{\psi_1}+\epsilon_{\psi_2}$
appears in the both $\psi_1$ and $\psi_2$ equations with 
equal magnitude.
This simple way to model the effect of the reinjection of the
reflected field into the laser cavity is used here for convenience.
The present model is intended only to display the new qualitative
features that arise when more than one mode is relevant.

\begin{figure}
\centerline{\hbox{
\psfig{figure=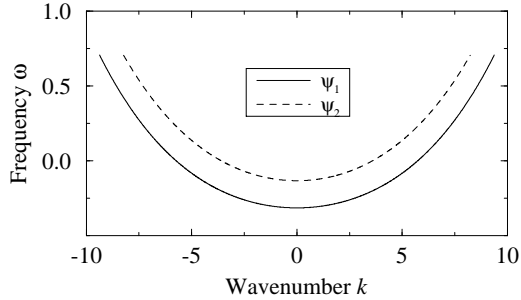,width=3.5in}
}}
 \caption{Dispersion curves for (unstable) traveling wave 
 	solutions in the two mode model.
	The solid (dashed) line represents solutions in 
	which only the favored (unfavored)
	mode is excited.
	The parameters are $\delta=.05$, $\sigma=.1$, 
	$\Omega=.001$, $a=.01$, $b=.01$,
	and $\alpha=-5$.
	}
 \label{fdisp}
\end{figure}

We are interested in the solution in which one longitudinal mode
supports a traveling wave and the other is inactive:
\begin{eqnarray} 
\psi_1 & = & \psi_k  \equiv  \rho \exp{i (kx-wt)}, \\
n & = & n_k  \equiv  \rho^2/b, \\
\psi_2 & = & 0, \\
\eta & = & 0,
\end{eqnarray} 
where
\begin{eqnarray}
\rho^2 & = & b \left[ r - 1 - \left(
	\frac{\Omega + \Delta- a k^2}{1+\sigma}
	\right)^2 \right] \\
\omega & = & \frac{\sigma \Omega + a k^2 - \Delta 
	+ \alpha \sigma \rho^2/b}{1+\sigma}.
\end{eqnarray}
The complementary solution is obtained by interchanging the
subscripts of the fields and taking 
$\Delta \rightarrow -\Delta$ in the expresions for 
$\rho$ and $\omega$.
Taking $\psi_1 = (1+B) \psi_k$, $\psi_2 = D \psi_k$, 
$n = (1+C) n_k$,
and $\eta = E$, we obtain the following linear equations for the
small fields $B,D,C,E$,
\begin{eqnarray} \label{linear2}
(\sigma+1) \partial_t B & = & -(2 a k+4 i a k 
{\tilde \sigma} \Omega
  - 4 i a^2 {\tilde \sigma} k^3) \nabla B \nonumber \\
 & &  + (i a - 2 a {\tilde \sigma} \Omega 
  + 6 {\tilde \sigma} a^2 k^2) \nabla^2 B \label{linear2B} \\
(\sigma+1) \partial_t D & = & \left[ i a \nabla^2 - 
2 i k - {\tilde \sigma} 
  (\Omega - \Delta + a \nabla^2)^2 \right. \nonumber \\
  & & \left. + i a k^2 + 
  {\tilde \sigma} 
  (\Omega + \Delta - a k^2)^2 \right] D \nonumber \\
 & & - \sigma (1+i \alpha) E^* \label{linear2D} \\
\partial_t C & = & b ( B + B^* - C) \label{linear2C} \\
\partial_t E & = & -2 b E + \rho^2 D^* \label{linear2E}
\end{eqnarray}

\begin{figure}
\centerline{\hbox{
\psfig{figure=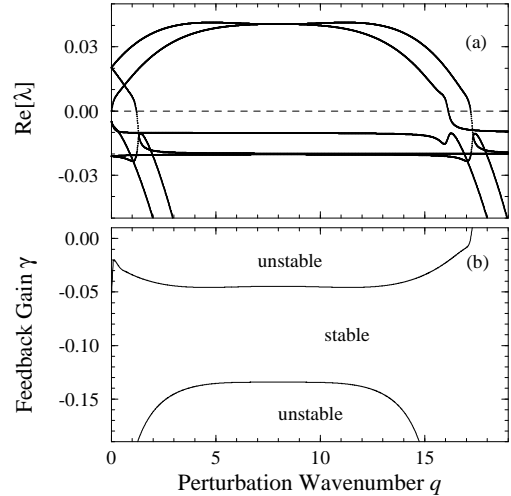,width=3.5in}
}}
 \caption{(a)  Real parts of the eigenvalues of an 
 	uncontrolled ($\gamma=0$)
	solution in which $\psi_2$ is a 
	traveling wave with $k=8$ and $r=1.5$, 
	and $\psi_1$ is zero everywhere.
	Parameters are the same as in 
	Fig.~4.
	(b)  Stable region of the same solution 
	with control with parameters
	$R=.5$, $\tau=2 \pi/\omega$, and $\Gamma=.25$.
	The traveling wave is stable at all 
	$q$ for $-0.13<\gamma <-0.05$.
	}
 \label{ftwomode}
\end{figure}

Fourier transforming, we again obtain a general expression
for the behavior of small differences of a 
perturbation wave number
from the controlled state.
Letting $q_x$ be the wave number in the transverse direction and
${\bf\xi} = (B, B^*, D, D^*, C, E, E^*)$, we have
\begin{equation}
\frac{d}{dt}{\bf\xi} = {\bf J} \cdot {\bf\xi} + 
{\bf M} \cdot \epsilon_{\bf\xi},
\label{general2}
\end{equation}
where here ${\bf J}$ is the matrix of coefficients 
obtained directly from Eqns.~(\ref{linear2B}-\ref{linear2E}) and 
\begin{equation}
 {\bf M} = \left(\begin{array}{ccccccc} 
			1 & 0 & 1 & 0 & 0 & 0 & 0 \\
			0 & 1 & 0 & 1 & 0 & 0 & 0 \\
			1 & 0 & 1 & 0 & 0 & 0 & 0 \\
			0 & 1 & 0 & 1 & 0 & 0 & 0 \\
			0 & 0 & 0 & 0 & 0 & 0 & 0 \\
			0 & 0 & 0 & 0 & 0 & 0 & 0 \\
			0 & 0 & 0 & 0 & 0 & 0 & 0 \\
		\end{array}\right).
\end{equation}
As in the case of the single longitudinal mode laser, 
a condition of the form
of Eqn.~\ref{def} defines the Floquet multipliers of the system.

\begin{figure}
\centerline{\hbox{
\psfig{figure=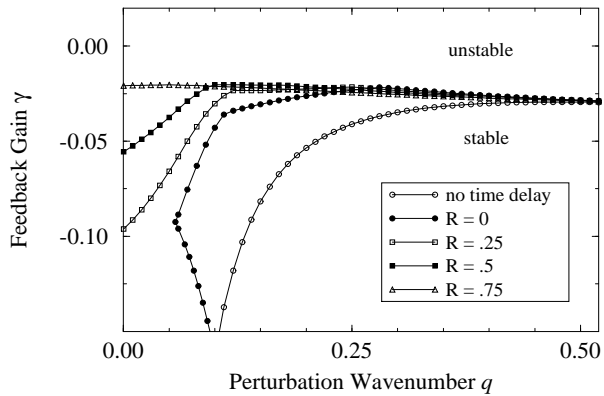,width=3.5in}
}}
 \caption{A detail from the left edge of the 
 	upper boundary of the
	domain of control in 
	Fig.~5(b) 
	shows the effect of
	varying $R$.  In the case of $R=0$, 
	as with no time delay,
	the traveling wave solution cannot 
	be controlled due to instabilities
	at small $q$.
	Larger values of $R$ do yield stable 
	solutions for sufficiently negative $\gamma$.
	}
 \label{ftwomodeR}
\end{figure}

We now describe the results of the linear stability
analysis of the two mode model.
As in the single mode model, each mode
is always unstable to transverse fluctuations,
but in the two mode model it is possible for one mode to
be unstable to the growth of the other as well.
A straightforward stability analysis of the 
{\em uncontrolled} equations shows that for all 
parameter choices both
$\psi_1$ or $\psi_2$ are marginally stable against transverse
fluctuations at $q_x=0$, but
only one of the modes is always stable against 
growth of the other mode.
Which mode is which depends upon
the choice of the mode separation, 
the desired wave number, 
and other parameters in the model.  
We will refer to a mode that is stable 
(unstable) against growth of other
longitudinal modes at $q_x=0$ as ``favored'' (``unfavored'').

The dispersion curves for transverse waves in the two modes have 
nearly the same functional form, but are
displaced relative to each other approximately 
by the mode spacing, $2 \Delta$.
(See, for example, Fig.~\ref{fdisp}.)
By choosing the wave number for the spatial filter,
one selects one traveling wave state from each of the two
dispersion curves.
Because the frequencies of these two states are different,
one can choose the time delay so as to suppress fluctuations
at the frequency of the undesired mode.
Thus it is plausible to suppose that 
the combination of the time delay and the spatial filter
capable of stabilizing either of the two longitudinal modes.
We will focus on the stabilization of an unfavored mode,
both because it would appear to be the more difficult case
and because it may be a better representation 
of the situation that arises in multimode systems.

Fig.~\ref{ftwomode} illustrates the stabilization of the 
unfavored mode ($\psi_1$) with the parameters 
listed in the caption.  
Fig.~\ref{ftwomode}a, shows the stability curves 
for the uncontrolled system,
clearly indicating the instability at $q=0$ that makes this case 
qualitatively different from the single mode case discussed above.
The spatial filter component in our control scheme is insensitive
to instabilities at or near $q=0$ because the filter must pass
components of both $\psi_1$ and $\psi_2$ with this wave number.
The width of the function $f(q)$ used for the spatial filter
will determine the range of $q$ which are passed.
As a result of the ineffectiveness of the spatial filter over
this range of perturbation wave number, the temporal component
of the control scheme must be relied on to 
stabilize these perturbations.

\begin{figure}
\centerline{\hbox{
\psfig{figure=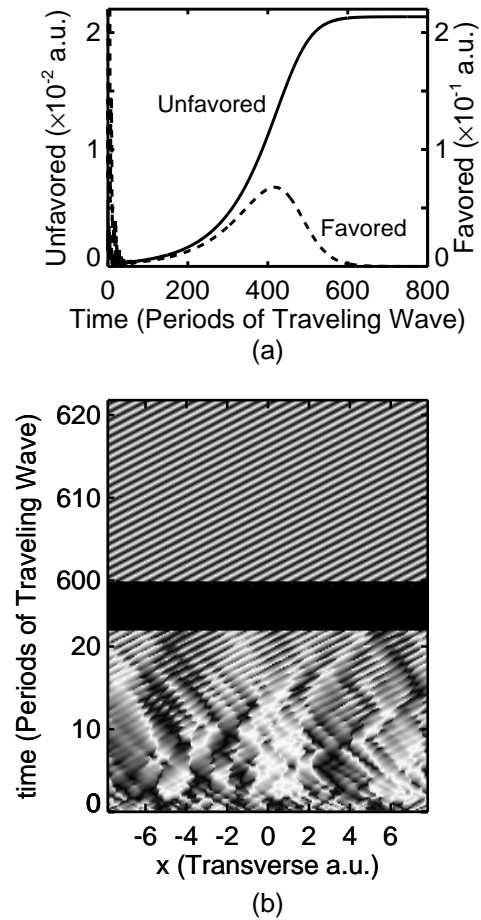,width=3.5in}
}}
 \caption{Evolution of two mode system with $a, b, 
 	\Omega, r, \sigma, \alpha,$
	 as in
	Fig.~4.
	with $\Delta=0.1$, $r=1.5$, $k=8$, $R=0.5$, 
	and $\gamma=.1$.
	(a) A spacetime plot of the phase of $\psi$ 
	in the controlled system.
	The lower region shows the dynamics 
	when the system is first turned on.
	After a transient time of approximately 
	200 periods of the desired orbit, 
	the system settles into the traveling wave state.
	(b) The magnitude of the favored mode (dashed line) 
	and unfavored mode (solid line) 
	as a function of time for the same run as shown in (b).
	Note the expanded scale at the right 
	for the favored mode.
	}
 \label{ftmevolution}
\end{figure}

The stability diagrams of Fig.~\ref{ftwomode}b and 
Fig.~\ref{ftwomodeR}
demonstrates that the time-delay
control is effective in controlling the range of 
perturbation wave numbers
that are not stabilized by the spatial filter.
Fig.~\ref{ftwomode}b shows that with both 
time-delay control (here with
$R=0.5$) and the spatial filter (with $\Gamma=0.25$) 
there is a range of
$\gamma$ that stabilizes the traveling wave solution.
Fig.~\ref{ftwomodeR} shows that when time-delay control is not
present, and also when $R$ is too small, there is no range
of $\gamma$ that stabilizes the traveling wave solution at all
wavenumbers.

A new feature that appears
in the two mode model, and is shown in Fig.~\ref{ftwomode}b,
is the lower boundary of the stable domain,
whose origin lies in the off-diagonal elements of ${\bf M}$.
When the system is not exactly on the desired orbit, there is
a finite amount of feedback generated.  
Because the desired mode has a much larger average magnitude 
than the other mode, the feedback signal is dominated 
by effects from the desired mode.  
This feedback is necessary to control the desired mode,
but it also affects the other mode.  
When the magnitude of this feedback becomes too large, 
as it must when $|\gamma|$ is increased, 
these unwanted perturbations to the undesired
mode cause the state to go unstable.

The position of the lower boundary of the domain of control
(Fig.~\ref{ftwomode}b) is important because it
determines the range of gain that can be used to obtain control.
If that range is very small, it may be difficult to find
an appropriate $\gamma$ in an experiment.  
Even worse, if the lower boundary becomes so high that
part of it reaches the lowest point of the upper boundary, 
there is no $\gamma$ which can control the system.
We find that the position of the lower boundary is affected
by several parameters.  
The lower boundary is raised when the pump rate $r$ is raised 
and when the wave number $k$ is lowered.  
The mode separation, $\Delta$, also plays an important role in the
location of the lower boundary.
For larger $\Delta$, the lower boundary is pushed down.
In a system in which $\gamma$ is the only adjustable
parameter ($r$, $\Delta$, and $R$ fixed), we find that
traveling waves with wavenumbers in a finite continuous
band can be stabilized.
The high-$k$ boundary of the band is determined by
the condition that traveling waves exist
(that $\rho$ must be real),
and the low-$k$ boundary is the point at which there
ceases to be a $\gamma$ that can control perturbations 
at all wave numbers.

As in the single mode model, numerical simulations confirm the
predictions of the linear analysis and show that the traveling
wave state can be obtained starting from a distant initial
condition.
Fig.~\ref{ftmevolution} shows the emergence of the desired
traveling wave from a low amplitude, noisy initial condition.
After an initial transient, the system clearly settles into
the desired pure traveling wave.

We have also observed the behavior of the system when $R$ is
chosen too small.
Although the only unstable modes in this case are very close to
$q=0$, we find that their growth completely destroys 
the traveling wave.
The system does not merely develop long wavelength modulations
of the desired wave.
We therefore conclude that both the temporal and spatial aspects
of the feedback signal we have analyzed play essential roles in
the success of the scheme.

\section{Conclusions}

Our study of the dynamics of laser Swift-Hohenberg equations
with time-delayed, spatially filtered feedback strongly suggests
that stable lasing at a single transverse wave number in wide
aperture lasers is possible.
A future publication will report on studies of a more realistic
model of field and carrier dynamics in a semiconductor system
of the type shown in Fig.~\ref{fsetup}, where preliminary
results are encouraging.
Though there are several nontrivial experimental issues associated
with the fabrication of such a device, we believe that this
is a promising direction for research and development.

We have presented a theoretical approach to the analysis of this
sort of feedback that appears to capture the relevant features
of the dynamics.
The linear stability analysis presented here is a straightforward
extension of some of previous work on stabilizing traveling waves
in the complex Ginzburg-Landau equation.
In the present case, however, the desired state seems to be
a global attractor, which gives us considerably more confidence
in its potential for practical implementation.

Finally, we would like to emphasize that the general method of
applying time-delay feedback combined with spatial filtering
is a powerful technique that might be adapted to many other
types of physical systems.
Its primary advantage is that the desired traveling wave state
need not be available in some external form for construction of
the feedback signal.
It is particularly suitable for optical systems, however, where
the necessary manipulations of the signal can be performed with
standard optical elements.

We thank Dan Gauthier and Rob Indik for helpful conversations.
MEB gratefully acknowledges the hospitality of the Arizona
Center for Mathematical Sciences.
JVM and DH were supported by AFOSR-94-1-0144DEF and AFOSR-94-1-0463.
JESS and MEB were supported by NSF Grant DMR-94-12416.

\end{multicols}


\begin{references}
\bibitem{modesep} The small mode separation, $2 \Delta$ is 
	assumed to have 
	a magnitude which is first order in the 
	expansion of Ref.~\cite{LMN2}.
\end{references}
\end{document}